# CAN URBAN AIR MOBILITY BECOME REALITY? OPPORTUNITIES, CHALLENGES AND SELECTED RESEARCH RESULTS

Henry Pak[1] (https://orcid.org/0000-0002-6259-3441),
Lukas Asmer[1] (https://orcid.org/0000-0002-5975-5630),
Petra Kokus[1] (https://orcid.org/0000-0002-2132-9572),
Bianca I. Schuchardt[2] (https://orcid.org/0000-0003-1251-3594),
Albert End[5] (https://orcid.org/0000-0002-9837-9762),
Frank Meller[3],
Karolin Schweiger[2] (https://orcid.org/0000-0002-8498-9535),
Christoph Torens[4] ( https://orcid.org/0000-0002-0651-4390),
Carolina Barzantny[5] (https://orcid.org/0009-0000-4672-1813)
Dennis Becker[6],
Johannes Maria Ernst[2] (https://orcid.org/0000-0001-8238-3671),
Florian Jäger[4] (https://orcid.org/0009-0002-9452-2792),
Tim Laudien[2] (https://orcid.org/0009-0005-9178-115X),
Nabih Naeem[3] (https://orcid.org/0000-0002-3144-3045)
Anne Papenfuß[2] (https://orcid.org/0000-0002-0686-7006),
Jan Pertz[1] (https://orcid.org/0000-0001-7638-5347),
Prajwal Shiva Prakasha[3] (https://orcid.org/0000-0001-5694-5538),
Patrick Ratei[3] (https://orcid.org/0000-0002-5161-8025),
Fabian Reimer[3] (https://orcid.org/0000-0001-6666-6903),
Patrick Sieb[7] (https://orcid.org/0000-0003-3316-5429),
Chen Zhu[6] (https://orcid.org/0000-0002-4320-4826),

With contributions by:
Rabeb Abdellaoui[2], Richard-Gregor Becker[9], Oliver Bertram[4], Aditya Devta[2], Thomas Gerz[10], Roman Jaksche[1], Andreas König[8], Helge Lenz[2], Isabel C. Metz[2], Fares Naser[2], Sebastian Schier-Morgenthal[2], Maria Stolz[2], Majed Swaid[1], Andreas Volkert[2], Kristin Wendt[8]

German Aerospace Center (DLR), Germany
[1]Institute of Air Transport, Köln
[2]Institute of Flight Guidance, Braunschweig
[3]Institute of System Architectures in Aeronautics, Hamburg
[4]Institute of Flight Systems, Braunschweig
[5]Institute of Aerospace Medicine, Hamburg
[6]Institute of Communications and Navigation, Wessling
[7]Institute of Maintenance, Repair and Overhaul, Hamburg
[8]National Experimental Test Center of Unmanned Aircraft Systems, Cochstedt
[9]Institute of Propulsion Technology, Köln
[10]Institute of Atmospheric Physics, Oberpfaffenhofen

Contact:
henry.pak@dlr.de
bianca.schuchardt@dlr.de

## Abstract

Urban Air Mobility (UAM) is a new air transportation system for passengers and cargo in urban environments, enabled by new technologies and integrated into multimodal transportation systems. The vision of UAM comprises the mass use in urban and suburban environments, complementing existing transportation systems and contributing to the decarbonization of the transport sector. Initial attempts to create a market for urban air transportation in the last century failed due to lack of profitability and community acceptance. Technological advances in numerous fields over the past few decades have led to a renewed interest in urban air transportation. UAM is expected to benefit users and to also have a positive impact on the economy by creating new markets and employment opportunities for manufacturing and operation of UAM vehicles and the construction of related ground infrastructure. However, there are also concerns about noise, safety and security, privacy and environmental impacts. Therefore, the UAM system needs to be designed carefully to become safe, affordable, accessible, environmentally friendly, economically viable and thus sustainable. This paper provides an overview of selected key research topics related to UAM and how the German Aerospace Center (DLR) contributed to this research in the project "HorizonUAM - Urban Air Mobility Research at the German Aerospace Center (DLR)". Selected research results that support the realization of the UAM vision are briefly presented.

## Keywords

Urban air mobility, air taxi, vertidrome, system-of-systems, market development, social acceptance

## LIST OF ABBREVIATIONS

| | |
|---|---|
| AAM | Advanced Air Mobility |
| ATC | Air Traffic Control |
| EASA | European Union Aviation Safety Agency |
| eVTOL | electrical Vertical Take-Off and Landing |
| FAA | Federal Aviation Administration |
| FATO | Final Approach and Take-off |
| IAM | Innovative Air Mobility |
| IFAR | International Forum for Aviation Research |
| MaaS | Mobility as a Service |
| MRO | Maintenance, Repair, Overhaul |
| NASA | National Aeronautics and Space Administration |
| SoS | System of systems |
| STOL | Short Take-Off and Landing |
| UAM | Urban Air Mobility |
| UAV | Unmanned Aerial Vehicle |
| VFR | Visual Flight Rules |

## LIST OF IMAGES

Fig. 1   Tiltrotor air taxi configuration
Fig. 2   Use Cases for Urban Air Mobility
Fig. 3   Exemplary vertidrome layout
Fig. 4   Social acceptance resulting from balancing the costs and benefits of a transportation system
Fig. 5   Central aspects of the HorizonUAM Project
Fig. 6   Capacity fade for tiltrotor aircraft with enhanced technology
Fig. 7   Payload-range-diagram for a degrading battery

## 1. INTRODUCTION

Urban Air Mobility (UAM) is a new air transportation system for passengers and cargo in urban environments, enabled by new technologies and integrated into multimodal transportation systems [1-3]. The transportation task is performed by electric or hybrid electric aircraft that take off and land vertically, remotely piloted or with a pilot on board. The aircraft used range from small drones for parcel deliveries to air taxis for passenger transport. While the term UAM is widely used in research and media, new terms are evolving such as Advanced Air Mobility (AAM) [4] or Innovative Air Mobility (IAM) [5]. For consistency, the term UAM will be used throughout this paper in the narrow sense of urban passenger transport only, unless otherwise noted.

The vision of UAM comprises the mass use in urban and suburban environments, complementing existing transportation systems and contributing to the decarbonization of the transportation system [6]. Users will benefit from time savings, and if battery electric propulsion systems are used, local emissions from UAM could be close to zero [1]. Safety, security, sustainability, privacy, and affordability are other features of UAM. UAM could thus help to maintain or even improve the quality of life in metropolitan areas despite the ongoing urbanization which is increasing the pressure on urban transportation systems and posing the challenge to large cities and metropolitan areas to ensure an efficient organization of transport [7].

Technological advances in many fields in recent decades have led to a renewed interest in urban air transportation. In particular, new opportunities have opened up for small aerial vehicles, which form the basis for innovative UAM concepts that provide fast and reliable transportation within cities, between suburbs and cities or between points of interest [8, 9]. The European Union Aviation Safety Agency (EASA) expects first commercial flights to take place in 2025 [1]. Furthermore, there are already plans to use these novel aircraft to transport passengers during the 2024 Olympic Games in Paris [10].

In recent years, several research and development activities have been conducted by research institutions, manufacturers and potential UAM operators with the aim of integrating these concepts into existing transportation systems in a way that is both economically viable and socially acceptable. Against the background of persistent challenges of current urban transportation around the world such as congestion, noise, and emissions there are numerous efforts underway to make urban air mobility a part of the overall urban transportation system.

The aim of this paper is to provide an overview of selected key research topics related to UAM and how the German Aerospace Center (DLR) is contributing to this research by combining its competencies in the areas of UAM vehicles, related infrastructure, operation of UAM services, and public acceptance of future urban air transport in the project "HorizonUAM - Urban Air Mobility Research at the German Aerospace Center (DLR)". For this purpose, the technical approaches and research results of the HorizonUAM project are briefly summarized and evaluated with respect to the goal of enabling an economically viable and socially acceptable UAM system.

The paper is structured as follows. Section 2 is dedicated to the visions of UAM and the expected opportunities for potential stakeholders. For this purpose, first the historical development of UAM is considered in order to derive how technological progress should enable a new era of urban air mobility. In Section 3, challenges of implementing UAM as a new part of the transportation system are discussed. In Section 4, the key research results of the HorizonUAM project are briefly presented. Section 5 summarizes the insights to date and concludes with their implications for a future urban aerial transportation system. Finally, it provides an outlook on the need for further research.

## 2. WHAT TO EXPECT FROM URBAN AIR MOBILITY: OPPORTUNITIES AND VISIONS

The advent of turbine-powered helicopters marked the beginning of a first wave of Urban Air Mobility between the 1950s and 1980s. In the United States, helicopters were used to transport passengers between major airports and downtown business districts in Los Angeles, Chicago, New York, and San Francisco / Oakland [11]. During this period there was a trend towards the use of larger helicopters. Passenger service began in the 1950s with a seven-seat version of the Sikorsky S-55. This was followed by the Sikorsky S-58 and Vertol 44-B, with capacities for 12 and 15 passengers, respectively. In the 1960s, the Boeing Vertol 107 and Sikorsky S-61L entered service with capacity ranging from 25 to 28 passengers.

Similarly, in Europe there were regular international helicopter feeder services to Brussels Airport from Lille (France), Amsterdam and Eindhoven (Netherlands), Cologne-Bonn, Dortmund and Düsseldorf (Germany) [12]. There was also a regular helicopter service connecting London Gatwick and Heathrow airports between 1978 and 1986 in order to avoid time-consuming transfers for occasional onward flights, which was discontinued with the opening of a freeway [13].

A completely different kind of helicopter commuter service was operated by Air General in the Boston area in the 1960s [11]. Within an 18-mile radius of Boston Logan International Airport, the carrier established about 70 helistops during its lifetime with about 40 of them in operation simultaneously. These helistops were privately owned, their use limited to Air General's helicopters, and generally located in motel parking lots, in the industrial parks, adjacent to businesses or on corporate campuses. The services were directly related to business travelers. The carrier had a fixed schedule, but flights were only operated if a reservation was made at least 30 minutes prior to the scheduled pickup time. Air General's fleet consisted of Bell Model 47-J2 and Model 47-G4a with three seats. Later in 1968, four-seat Jet Rangers were placed in service.

These initial attempts to create a market for urban air transportation did not remain permanently established. This was due to a lack of profitability and of community acceptance [11, 14, 15]. Operators were forced to cease operations due to the removal of helicopter subsidies by governments and major airlines, increasing public protests against aircraft noise, and safety concerns after several accidents occurred.

As a result, passenger transportation by helicopter today exists only as a niche market in a few metropolitan areas, such as Los Angeles, New York, and Sao Paulo. Charter flights for companies and wealthy individuals as well as sightseeing flights for tourists are the main services offered [15]. In addition, there are helicopter flights for medical purposes, such as rescue missions or transporting patients to and from hospitals in urban areas.

Currently, the helicopter services offered by BLADE Urban Air Mobility probably come closest to the vision of UAM as a means of transport accessible for a broader community. In addition to charter flights, BLADE offers regularly scheduled short-haul flights [16]. Even though the company does not have a license to operate aircraft themselves, it acts as an intermediary between passengers and licensed air carriers. In New York, flights are offered between Manhattan and both John F. Kennedy (JFK) and Newark airports. In Canada, there are connections between downtown Vancouver and the cities of Victoria and Nanaimo, located on offshore Vancouver Island. In France, the Principality of Monaco is connected with Nice Airport via helicopter service. However, the prices for the flight are significantly higher than the prices for a regular taxi trip. For example, BLADE offers transfers between JFK Airport and Manhattan starting at US $195 for a single seat, while taxis charge a flat rate of US $70 or US $75 during peak hours for trips of up to four people [17]. The high prices are one reason why passenger transport between the airport and the city by helicopter has been a niche activity.

The ever-increasing pressure on transportation systems in megacities and metropolitan areas results in undesired delay as congestion lengthens travel times in urban areas. Not only does this mean a loss of comfort for users, but the environment is also burdened by additional emissions. The trend towards urbanization will further increase the pressure on transportation systems by steadily increasing trip numbers and distances travelled due to the growth of urban areas. The search for alternative transportation options through novel technologies is on the rise.

Technological advances in numerous fields over the past few decades have led to a renewed interest in urban air transportation. These technologies include energy storage, automation and artificial intelligence, sensors, communications and navigation, as well as simplified ridesharing and trip booking enabled by the proliferation of smartphones [15, 18]. For small aircraft in particular, new opportunities opened up with novel aircraft designs that do not automatically fall into any of the previously common categories. Different from conventional helicopters, these novel designs, which include multi-rotor, tiltwing, tiltrotor, and powered-wing concepts with vertical take-off and landing capabilities, utilize distributed propulsion. An exemplary tiltrotor air taxi configuration is shown in Fig. 1. Propulsion systems range from battery electric to hydrogen electric and hybrid to gas [2]. Considering the related design characteristics, these types of novel aircraft are often referred to as (electric) Vertical Take-off and Landing vehicles ((e)VTOL). Although initial operations are expected to be conducted with a pilot on board, such operations are expected to be remotely piloted, automatic and eventually fully autonomous in the future [6, 19]. The aircraft currently being developed for use in urban transport, are expected to be, among other things, small and less noisy than conventional helicopters [20], making continuous and high-density operations in urban areas more conceivable.

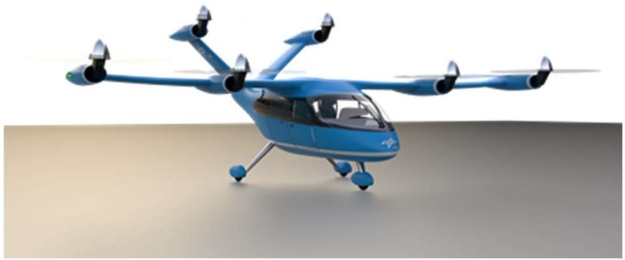

**Fig. 1** Tiltrotor air taxi configuration

Based on this new category of aircraft [21], the concept of Urban Air Mobility is gaining new momentum. Several use cases for urban passenger transport by air have been proposed, with air taxis, airport shuttles, air metro services, and intercity services (Fig. 2) being the most promising forms [19, 22, 23]. Air taxi services are passenger flights performed by small aircraft on demand between all available landing sites within a defined area [3, 9, 15, 22, 23]. Occasionally, on-demand air mobility is also referred to in this context. Airport shuttle services are passenger flights between various points in the city and an airport. The flights are usually operated according to a fixed schedule [3]. It is assumed that airport shuttle services will evolve as first form of UAM, initially for first-class airline passengers and later for all air travelers to and from the airport [22]. The Air Metro service is similar to today's public transportation systems such as subways and buses, and can operate on predetermined routes with fixed stops and according to regular schedules in busy areas of the city [23]. In particular, this service is a transportation alternative for people commuting between suburbs or satellite cities and the city center. Intercity services are passenger flights between two cities that are more distant from each other. In this case, the transport distances are typically larger so that the use of Short Take-off and Landing (STOL) aircraft instead of VTOL aircraft [19] is an additional option. Thus, this use case is quite similar to Regional Air Mobility, which operates from airfields with a runway and with aircraft larger than those used in urban areas.

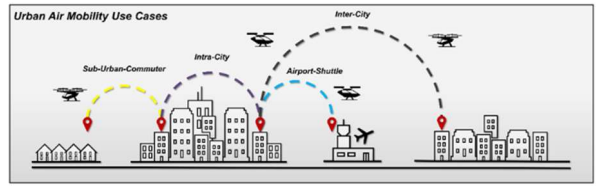

**Fig. 2** Use Cases for Urban Air Mobility [19]

In order to provide new air mobility for passengers and cargo within or around cities and regions, novel landing sites suitable for UAM operations are required. So-called vertidromes range in size and functionality: small vertistops with only one single final approach and take-off area (FATO); medium-sized vertiports with a few FATOs, charging/fueling systems and facilities for minor maintenance repair and overhaul (MRO) operations; large vertiports that can accommodate dozens of FATOs, MRO infrastructure, parking spaces for aircraft and possibly an operations control system or office space for staff [24-27]. An exemplary vertidrome layout is shown in Figure 3. They can be located on the ground or on the roof of parking lots, train stations, or other suitable buildings. A large number of vertidromes located close to centers of demand or well connected to transit and the road network ensures that many people can rapidly reach the take-off vertidrome from their origin and then quickly reach their final destination from the destination vertidrome. Thus, more customers are attracted to use UAM services leading to growing demand with positive operational feedbacks such as higher aircraft utilization as well as lower per-vehicle cost due to high aircraft production volumes [8]. Integrating UAM with Mobility-as-a-Service (MaaS), which aims to provide users with a range of mobility services tailored to their needs, could also help increase the utilization of the UAM system. Ultimately, as the number of users increases, prices for all users could decline.

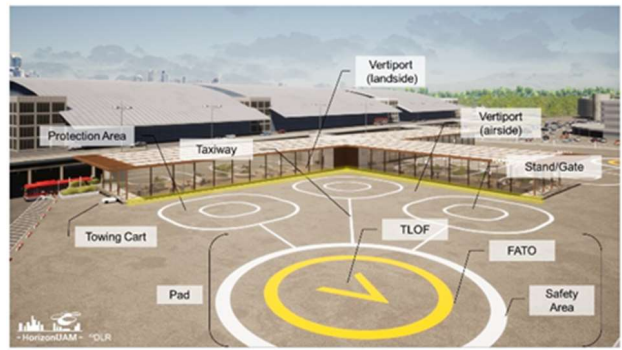

**Fig. 3** Exemplary vertidrome layout [24]

Urban air mobility has the potential to provide advantages and benefits to multiple stakeholders and in different dimensions: Key benefits for users can be shorter travel times at fares that in the long run can be significantly lower than those of today's helicopter services [22]. Time savings result from direct flights in cities and the high speed of the aircraft. The time advantage will become even more important when alternative ground transportation is disrupted during rush hours [28]. In addition, UAM can offer shorter travel times on second-tier connections, where demand is not high enough to establish high-quality ground transportation. When combined with MaaS, offering transportation services from multiple transportation providers through a unified gateway, users can benefit from seamless and comfortable transportation.

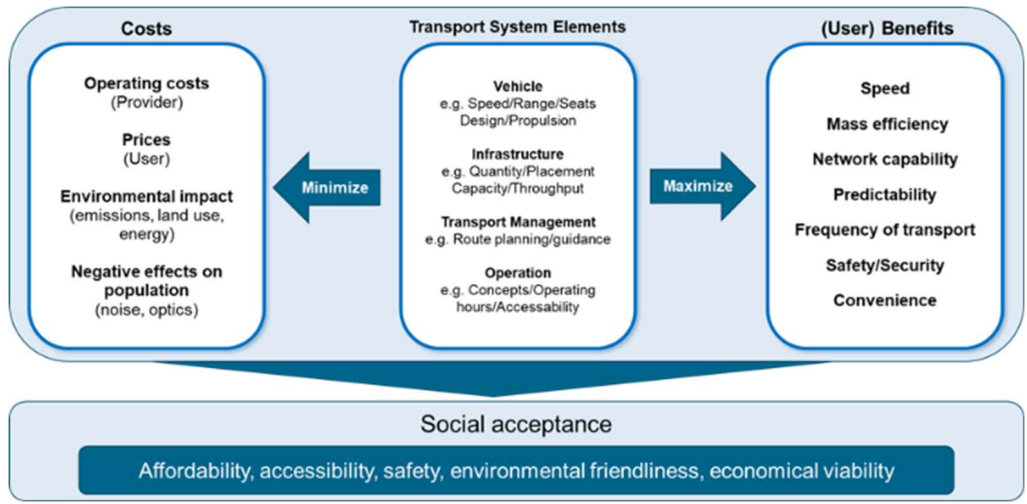

**Fig. 4** Social acceptance resulting from balancing the costs and benefits of a transportation system

UAM is also expected to have a positive impact on the economy by creating new markets and employment opportunities for manufacturing and operation of UAM vehicles and the construction of related ground infrastructure. Recent studies predict that up to 5 million UAM vehicles could be in operation worldwide by 2050, most of them for the inner-city use case [29]. Since most UAM vehicle concepts rely on electric propulsions, local emissions from UAM in urban environment could be close to zero [1].

As UAM can benefit the users, the economy and the environment, its development is being promoted by many institutions. In particular, the European Commission describes UAM as a system that effectively complements the existing transportation system, contributes to decarbonization, and provides benefits to citizens and communities in its Drone Strategy 2.0 [6]. With this strategy, the European Commission aims at promoting the development of the drone ecosystem, including UAM, and contributing to the realization of the vision of UAM. In parallel, aviation authorities such as the European Union Aviation Safety Agency (EASA) or the Federal Aviation Administration (FAA) are working on a regulatory framework to enable the deployment and implementation of UAM operational concepts [4, 30, 31].

## 3. CHALLENGES TO URBAN AIR MOBILITY BECOMING A REALITY

Like any transportation system, UAM is a complex system of systems, which is characterized by components (constituent systems) that interact in a variety of ways. Changes in individual components can influence each other in either positive or negative ways, reinforcing or opposing each other. As a change in one component of a system can have an impact on other components and on the behavior of the system in general, the system performance results from the interaction of its components with impact on relevant stakeholders such as users, operators, and society in general as well as on the environment.

In order to create a viable urban transportation system, it is necessary to harmonize the individual system components such as vehicles, infrastructure, and the operational framework including the air traffic management (ATM), or more precisely Unmanned Aircraft System Air Traffic Management (UTM) in the context of UAM. In order to balance the interests of all relevant stakeholders, it is essential to compare and reconcile costs and benefits when designing and evaluating transportation systems (Fig. 4). By systematically comparing benefits and costs, the efficiency, sustainability, and societal impact of different configurations of system components can be evaluated.

The evaluation of benefits can include different aspects such as travel speed, mass efficiency, network capacity, predictability and frequency of transport services, safety/security and convenience. Costs include monetary costs, such as fares for users, as well as operating costs and investments to build and operate, for example, the transport infrastructure. They also consider external costs, such as environmental and social impacts on residents and communities, like noise pollution. It is critical to ensure a comprehensive and realistic assessment of the new transportation mode by considering these external costs in the evaluation. By minimizing costs and maximizing benefits, a sustainable transportation system is created that meets society's expectations and requirements.

In contrast to existing transport modes, UAM is still at the concept stage. This means that there are many options for designing the elements of the transportation system to minimize costs and maximize benefits. The challenge is the harmonization of the system elements in such a way that the resulting UAM transportation system is socially accepted.

**Social acceptance** is seen as the key to a successful implementation of UAM [6]. Although the term is often used, clear definitions of social acceptance are rarely given. A concept of social acceptance originally developed in the field of renewable energy can be used as orientation. According to [32], three dimensions of social acceptance can be distinguished: socio-political acceptance, community acceptance, and market acceptance. Socio-political acceptance refers to the social acceptance of technologies and policies by the public, key stakeholders, and policy makers at the broadest and most general level. In contrast, the specific acceptance by local stakeholders, such as potential users, non-users and local authorities, is referred to as community acceptance. The third dimension of social acceptance is market acceptance, which is evidenced by the adoption of innovative products by consumers and investors' acceptance. These three categories of social acceptance can be interdependent and correspond to different stakeholders.

In the field of UAM, there is a wide variety of stakeholders such as potential users, the UAM industry, governments, public institutions, regulators and indirectly affected third parties, which are important for social acceptance, each with their own motivations, expectations and concerns regarding UAM [1, 2]. Potential users include urban residents, commuters or travelers who expect UAM to save time and provide safe, reliable, convenient and affordable transport from origin to destination. They may be concerned about noise, safety and security, and environmental impact. All entities directly involved in the design, manufacture, operation and maintenance of UAM aircraft and services form the UAM industry and are motivated to make a profit from their activities. This stakeholder group is interested in a stable regulatory framework, minimal bureaucracy, support for the development of a new industry, access to a skilled workforce and favorable taxation. The impact of regulation on the economics of UAM, excessive regulation, public opinion, nimbyism and environmental issues are some of their concerns. Governments, public institutions and regulators are a subset of stakeholders that include bodies and authorities at supranational, national and local levels. Their motivation focuses on public welfare, public safety, an efficient mobility system, limiting congestion and pollution, creating jobs, supporting and building an industry, the environment and public opinion. Their expectations of UAM are positive contributions to the community, income tax and that the industry complies with regulations. This stakeholder group has concerns about public opinion, loss of life, impact on voters, prestige for their respective jurisdictions, under- or over-regulation, and environmental issues. Furthermore, there is an indifferent group of indirectly affected third parties, including private individuals, professionals, associations, extended industry, potential competitors and non-users of UAM. These stakeholders will most likely evaluate UAM based on how it benefits themselves and/or society. In particular, people skeptical of UAM may be concerned of safety, privacy, noise and aesthetics [2].

For UAM to be accepted by society at large, the societal and environmental impacts should be addressed in a way that promotes benefits and minimizes risks, so that UAM as part of the future urban transportation system would be safe, affordable, accessible, and environmentally friendly and thus sustainable [2, 6]. Achieving these characteristics places requirements on the components of the UAM system and their interactions that can be challenging to meet. Further guidance on the development of a sustainability framework for UAM in the context of sustainable urban mobility planning is provided by [33, 34], which give a detailed description of desired characteristics associated with the demand for sustainability. Some of the challenges and their interactions are discussed below.

Concerns about **safety** may be an initial barrier to the adoption of UAM. There are fears that UAM users, other airspace users, and persons on the ground may be endangered. Novel aircraft designs, types of propulsion, and concepts of operation in an urban environment require the evolution of the regulatory environment to ensure safety and thus drive public acceptance of UAM. EASA has started the regulation process with setting up the new aircraft category VTOL [21], implementing regulations on U-space [35], and prototyping vertidrome specifications [36]. UAM will need to fit into the existing aviation system which has been established and is supported by a large number of regulations and standards. Integrating UAM into this system is expected to be a major challenge for many years to come [37].

Key areas for regulation are [2, 38]:

- Manufacturing, operation, and maintenance of aircraft, especially for autonomous and aircraft without pilot on board.
- Certification of pilots, aircrew, maintenance, and other personnel.
- Certification of aviation facilities, e.g. vertidromes.
- Operation of a network of facilities for air navigation, airspace and (unmanned aircraft system) air traffic management, especially for low altitudes.
- Safe, secure and efficient integration of air taxis into the airspace alongside with manned aviation [35].
- Global harmonization of UAM air navigation procedures facilitating cross-border UAM operations.

There is a need to amend the existing rules for airworthiness, air operations, flight crew licensing and Rules of the Air [30, 31]. The amendment of existing regulations or the creation of new ones can be of a fundamental nature. For example, the existing SERA (Standardized European Rules of the Air) provisions imply that current aircraft operations in an urban area may be performed for a very specific purpose, mainly police helicopters and helicopter emergency medical services [30]. Therefore, changes to SERA are necessary to enable the future transportation of people over densely populated areas with a level of safety that is at least as high as for operations with conventional airplanes or helicopters.

With regard to mixed operations of manned and unmanned aircraft, implications for the preferred routing scheme may occur in order to meet the goal of safety [30]. Predefined routes allow for systematic deconfliction between aircraft, thus automatically avoiding mid-air collisions. Free routing would require reliable detect and avoid capabilities among unmanned aerial systems, which are under development but not yet certified. If in the future safety can be guaranteed without the need for predefined routes or areas/corridors, then this potential limitation could be removed with positive implications for flexibility, and thus the operating costs and travel times.

With respect to the requirement of a trustworthy UAM system, not only safety but also security is an essential aspect. While safety is concerned with the prevention of accidents and technical failures, security is concerned with the protection against criminal or harmful acts from the outside. A distinction can be made between physical threats and cyber threats [39]. In the context of safe UAM operations, a large amount of data needs to be exchanged, with a particular focus on ground-to-air and air-to-air communications. Artificial intelligence and machine learning as well as their certification also play an important part of the further development of UAM and, therefore has to be evaluated from a cybersecurity perspective [40]. Against this backdrop, these systems must be appropriately secured to protect them from external attacks. In order to achieve this, operational concepts must be developed and harmonized from the outset with cybersecurity in mind in order to enable safe and secure operations [41].

New regulations may also be required as operations in urban environment may present hazards that are different to or more pronounced than those of conventional aviation in higher altitudes. For example, adverse weather conditions could pose challenges regarding safety, but also reliability of UAM [2, 22, 42]. Extensive urban development with tall buildings impedes the air flow and leads to locally increased turbulence, gusts, channeled flows or blockages with flow diversions and vortex shedding, and vertical wind systems. Therefore, to ensure safe operations in an urban environment, even in moderate weather conditions, additional measures are required. For example, aircraft and vertidromes should be equipped with appropriate meteorological infrastructure, such as high-resolution wind measurement [43], which may have a negative effect on operating costs.

**Affordability** refers to costs relative to income and, therefore, people's ability to pay for UAM services. In particular, it considers how transportation systems affect the mobility of individuals with lower incomes compared to people with middle or high incomes. There are concerns that UAM may not be affordable for lower- and middle-income households [44]. As part of social acceptance, UAM services should not be limited to the "wealthy few" [6]. However, lack of affordability, and thus lack of social equity, may be a major barrier to social acceptance.

Since operating costs and ticket prices are closely linked, the challenge is to reduce operating costs. Failure to achieve significantly lower operating costs compared to helicopters will jeopardize the affordability and thus market penetration. At present, the costs for transporting passengers with air taxis and thus UAM ticket prices are still highly uncertain [37]. While proponents expect ticket prices to become comparable to those of conventional taxis or even lower than the costs of using a privately owned car [8], the following analyses come to different conclusions. Booz Allen Hamilton [22] estimated a passenger price per mile of between US $6.25 (3.75€/km) and US $11 (6.60€/km) in the first few years of operation, depending on the size of the aircraft, compared to about US $3.00 (1.80€/km) for the conventional taxi and well below US $1.00 (0.60€/km) for using a private car. Pertz et al. [45] found that short-distance missions within a city, exemplified by intra-city and airport shuttle use cases, result in a fare per kilometer of 4€ to 8€. Regional missions with a longer mission distance can decrease this value to as low as 1€ per kilometer. A high load factor, as well as a certain number of flight cycles per day, are presupposed.

There are many ways to reduce operating costs such as reducing aircraft manufacturing costs, increasing energy efficiency, reducing maintenance and repair costs or cycles, reducing costs to build and operate vertidromes, and improving network efficiency through high utilization and load factors. As often mentioned, automation can play a key role in reducing operating costs by reducing pilot training requirements and eventually not to rely on a pilot on-board at all [46]. At the same time, this leads to technological and regulatory challenges [6, 47].

Furthermore, potential cost reductions will also depend on operators achieving economies of scale [37]. These economies of scale will depend on the specific business model of the UAM operator and on the regulatory landscape that defines operator costs – such as potential limitations on landing and take-off infrastructure, restrictions on access to airspace, regulatory requirements that impose costs on operators, or potential public subsidies for the provision of UAM services.

**Accessibility** in general refers to the ease with which individuals can reach destinations in order to participate in activities or obtain desired services. In the context of UAM, accessibility refers to the ability to quickly reach the departure vertidrome from the origin of a trip and the destination of a trip from the arrival vertidrome. Analyses have shown, that access and egress times to and from the vertidromes (as well as processing times to change modes) strongly influence the attractiveness and thus the demand for UAM [8, 9]. Accessibility can be improved by locating the vertidromes closer to centers of high demand or by reducing travel time to the vertidromes through intermodal connections to transit and the road network [48]. The placement of vertidromes is an optimization task: a balance must be found between site requirements, costs and other issues, such as nuisance to neighbors from noise (see below) and visual pollution to avoid threating social acceptance [6]. Vertidrome placement is also critical to mission composition in terms of flight distances and recharging capabilities. This has a major impact on the optimal aircraft design, fleet composition and the associated trade-off between hover/vertical flight efficiency and cruise efficiency [38], and thus operating costs.

In addition to optimal placement, the right vertidrome throughput capacity is necessary to support the development of UAM. Vertidromes could be one of the most significant capacity limiting and cost driving aspects of UAM [47]. Depending on their structure and size, there is a wide range of vertidrome capacities in terms of aircraft movements and passenger throughput being suggested in the literature. Numbers range from less than 10 to over 1000 operations or up to 1400 passengers per hour [24, 49]. Three archetypes of vertidromes with different sizes and capacities are emerging: small vertistops, medium-sized vertiports, and large vertiports (Sect. 2). In addition, each facility is equipped with approximately twice as many parking spots. Size information varies widely. According to Johnston, RiedelSahdev [26], the footprint of a vertistop with two parking / charging spots is 20 by 30 meters with capital expenditures in the range of US $200,000 to US $400,000 and operating costs (without cost of power for charging or refueling) per year between US $600,000 and US $900,000. However, according to Taylor, SaldanliPark [50], the typical dimensions of a Multi-Function Single Pad without parking / charging spots are 39 by 69 meters. When separate staging areas are integrated, the footprint increases to 72 by 99 meters, nearly the size of a soccer field. Estimated costs range from US $350,000 to US $950,000. As the largest ground infrastructure, large vertiports are estimated to cover 120 by 55 meters and to cost US $6 million to US $7 million to build and US $15 million to US $17 million per year to operate [26]. To justify the land used and the investment and operating costs, there must be a minimum number of passengers and flights, requiring affordable UAM services. Otherwise, "the land would be better allocated as a park or community garden" [51].

The goal of **environmentally friendly mobility** is to maintain and ensure the mobility of people and goods without placing excessive burdens on people and the environment in terms of greenhouse gas emissions, air pollutants, noise, land use, wildlife and resource consumption [52]. While novel aircraft concepts are based on electric or hybrid-electric propulsion and, thus, contribute in combination with renewable energy sources to the decarbonization of the transportation system, noise is perceived in the literature or by experts as a predominant risk of UAM [1, 46]. This includes the noise generated by the vehicles during take-off, landing and flight. To be accepted, UAM aircraft must be significantly quieter than today's helicopters [3, 8, 20]. This presents a significant challenge to the aircraft design process. As described by Straubinger et al. [38], possible approaches to noise reduction include the distribution of thrust production to multiple rotors, minimizing take-off weight, and/or increasing the rotor area. The latter is in contrast to the requirement for air taxis to be as compact as possible to cope with the limited space in cities, and thus aircraft size and seat numbers could become the limiting factor for the throughput of vertidromes in urban areas.

In addition to reducing noise at the source through engine and airframe technologies, there are other ways to reduce noise: through operational measures such as low-noise procedures on ground and in the air and dedicated flight paths, through compatible land use and urban development, and through operational restrictions such as flight quotas, noise limits, or curfews [53]. The application of these measures can help to increase the acceptance of UAM with respect to noise. For example, predefined routes (see above) or no-fly-zones in a free-flight scenario would help to systematically avoid flying over areas and buildings that require noise protection. The same applies to visual pollution and protection of privacy. However, these measures make economic viability more difficult. Predefined routes may be longer, which increases costs, and flight quotas and curfews limit the number of flights that can be offered. As a result, affordability may suffer.

Based on the aspects of safety, affordability, accessibility and environmental friendliness, this chapter has discussed an excerpt of the challenges for a future UAM system. Since a UAM system is characterized by a high degree of complexity, changes in subcomponents usually result in changes in the entire system. Therefore, the overall challenge is to harmonize different stakeholder requirements and optimize the interactions between the individual components of the system in such a way that UAM is fully acceptable by society and successfully integrated into current transportation networks. As Straubinger points out, "successful UAM introduction relies on a broad variety of factors that have to be considered and still a large number of questions stay unaddressed in the existing literature" [38]. With the Horizon UAM project, DLR is contributing to the investigation of UAM. An overview of the results is presented in the following chapter.

## 4. ADDRESSING THE CHALLENGES: SELECTED RESULTS FROM DLR'S RESEARCH PROJECT HORIZONUAM

Research on Urban Air Mobility has been taking place at the German Aerospace Center in recent years focusing on individual fields and challenges. The project HorizonUAM ([54], Fig. 5) has now brought together ten DLR institutes from various fields to research on the vision of Urban Air Mobility holistically. For the first time, HorizonUAM combined the research about UAM vehicles, the corresponding ground infrastructure, the operation of UAM services, as well as the public acceptance of future urban air transportation. Competencies and current research topics including propulsion technologies, flight system technologies, communication and navigation went along in conjunction with the findings of modern flight guidance concepts and operational techniques at vertidromes and conventional airports. The project also analyzed possible UAM market scenarios up to the year 2050 and assessed economic aspects such as the degree of vehicle utilization and cost-benefit potential via an overall system model approach. Furthermore, the system design for future air taxis was carried out on the basis of vehicle family concepts, onboard systems, aspects of safety and security as well as the certification of autonomy functions. The analysis of flight guidance concepts and the scheduling, sequencing and traffic flow analysis of air taxis at vertidromes was another central part of the project. Selected concepts for flight guidance, communication and navigation technology were demonstrated with drones in a scaled urban scenario.

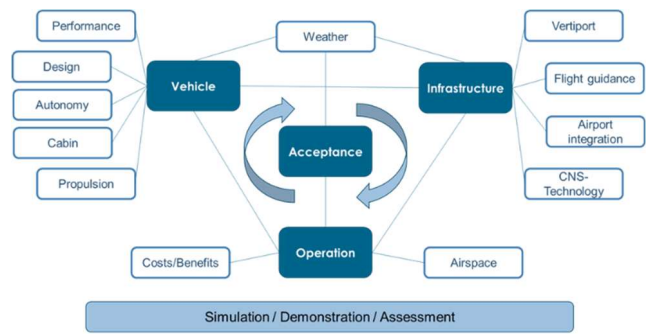

**Fig. 5** Central aspects of the HorizonUAM Project

In the following, selected research results are briefly presented that contribute to making the vision of UAM a reality. Particular, the focus will be on how the technical approaches and research results can contribute to minimizing costs, maximizing benefits, and increasing societal acceptance, considering the requirement that the UAM system needs to be safe, affordable, accessible, and environmentally friendly.

### 4.1. Vehicles

In addition to vehicle design, the HorizonUAM project included vehicle-related research on battery degradation, onboard systems, and cabin design.

#### 4.1.1 Vehicle design

The vision of UAM comprises comfortable, emission-free, fast, and safe air transportation in densely populated and congested urban centers. With (hybrid-)electric propulsion, autonomous flight, and low-noise vertical lift capabilities under development, the eVTOL vehicles required for air taxi services in the context of UAM certainly face several challenges at the system level alone. However, to facilitate the air taxi services within complex urban environments, many interdependencies between the vehicle and other stakeholders of the UAM ecosystem have to be considered in the early conceptual design phase in order to ensure smooth simultaneous development of the multiple system elements involved.

Therefore, a system-of-systems approach is adopted to explore such multidisciplinary problems with distributed, interacting system elements, where vehicle design becomes an integrated part of the overall UAM transportation system development. This makes it possible to assess impacts of the vehicle design on the overall system and vice versa to address challenges and opportunities of the UAM vision. The foundation of the system-of-systems approach for the vehicle design studies lies in the integration of agent-based simulations during the conceptual design phase. Employing a simulation toolkit tailored to UAM, the methodology supports standalone simulations of fleet operations [55]. Here, the focus is on vehicle and fleet design, while passenger demand, vertidromes, trajectories, etc., are only modeled in a simplified way or based on assumptions. The fleet simulations enable the assessment of metrics such as transport capacity, dispatch efficiency, energy demand, etc., in the fleet operational context of a UAM transportation system, as demonstrated in [56]. Additionally, a holistic vehicle design workflow is established, integrating aircraft architecture, cabin concepts, and onboard systems. This comprehensive approach allows for the design and assessment of UAM from the subsystem, system, and system-of-systems levels [57]. Life cycle assessments are also incorporated to evaluate the sustainability of battery technology and energy production [58].

Resulting from these studies, two vehicle concepts were developed to capture the design space of possible UAM vehicles and use cases: a wingless multirotor for urban missions and a winged tiltrotor vehicle for suburban or megacity missions. These concepts encompass specifications such as capacity, range, and speed based on the HorizonUAM use case definitions [19]. Both vehicles can carry four persons and are tailored to their respective vehicle performance characteristics. As such, the multirotor is sized for a range of 50 km at 120 km/h with two intermediate stops, and the tiltrotor is sized for a range of 100 km at 210 km/h with one intermediate stop. Both concepts are designed to withstand adverse weather conditions and hold minimum energy reserves for a 20-min loiter time. Fleet simulations demonstrate operational results, showcasing average flight hours of about 9.5 flight hours per day for each vehicle concept. In their respective transport network, the multirotor is operated on up to 40 missions with an average range of 22 km, and the tiltrotor performs up to 22 missions at an average range of 82 km. Furthermore, the sensitivities of propulsion technology, operational strategies, and fleet planning are considered. All details about the design process and vehicle concepts can be found in [59].

Making the vision of UAM a reality requires collaborative, multidisciplinary, and simultaneous development of all system elements. The introduced UAM system-of-systems approach has addressed this challenge and shed light upon several early design phase research questions. We have observed that early adaptations of UAM may be feasible, but are most likely constrained by piloted, low-capacity and short-range vehicles for small scale intra-city missions. Missions over longer ranges suitable for megacities open up the possibility of energy-efficient and time-saving air transport, but require advances over the current state of technology. Thus, continued efforts in (hybrid-)electric propulsion, autonomous flight, and low-noise vertical lift capabilities are needed to unfold UAM's full potential.

### 4.1.2 Battery degradation

The propulsion battery system is one of the main systems of the UAM and is subject to degradation during operation while it is expected to be a main cost driver. The battery ageing depends on the system design and the way the aircraft is operated. Replacing a degraded battery increases the life-cycle expenses of the vehicle and also consumes resources. Therefore, it is elementary to understand the impact of design and operational decisions on battery ageing. The degradation of batteries is investigated for the wingless multirotor for urban missions and the winged tiltrotor vehicle for suburban or megacity missions and their operation. The aircraft concepts are equipped with Sanyo UR18650E NMC-type lithium-ion batteries. Lithium-ion batteries are exposed to various ageing mechanisms, which can be found, for instance, in [60-62]. Considering battery lifespan from an operator perspective, there are three primary stress factors: temperature, state of charge, and the load profile.

An empirical degradation model is selected to simulate the capacity fade, as it offers an appropriate detail level coupled with swift runtimes. Schmalstieg et al. [63] present an empirical ageing model for lithium-ion batteries, that combines degradation caused by the cycled charge and the calendar ageing independent of the usage. Typically, the ageing of lithium-ion batteries is categorised into three phases: The strong initial degradation, a linear degradation, and the rapid capacity fade [64]. The model by Schmalstieg et al. reproduces the first two phases, but excludes the rapid capacity fade. The tipping into rapid fade for lithium-ion batteries is usually excepted at approx. 80% of the original capacity. However, the data by Schmalstieg et al. is provided down to 75% of the initial capacity and hence, the fading is considered until 75% of the initial capacity.

Our study's assumptions and limitations are following: A constant battery temperature of 26°C is assumed [65]. All battery cells are equal in terms of ageing. The empirical degradation model of 2014, reflects the battery technology of its publication year. The battery is charged with 1C after each flight mission to a state of charge of 100%.

The two aircraft concepts are a multirotor and tiltrotor aircraft, that are designed for a near and far term technology level. Different aircraft utilisations are researched, but independent of the utilisations, the degradation is primarily caused by the load cycles rather than the calendar aging. For the multirotor concept, 80% of the original capacity is reached after 1470 to 1610 cycles depending on the payload. For a safe operation under full payload only 55% battery capacity is required. Therefore, extending battery usage below 80% to a capacity fade of 75% could increase battery cycles to 2300 to 2500. Still, the batteries would have to be replaced every two to four months.

The tiltrotor concept was developed for longer distances and demands more energy to cover the design range and additional 20 minutes reserve. Consequently, the battery is designed that 79% of the original capacity is required. Even if the battery is used until a remaining capacity of 79% is reached, the lifetime reaches only 490 to 560 flight cycles.

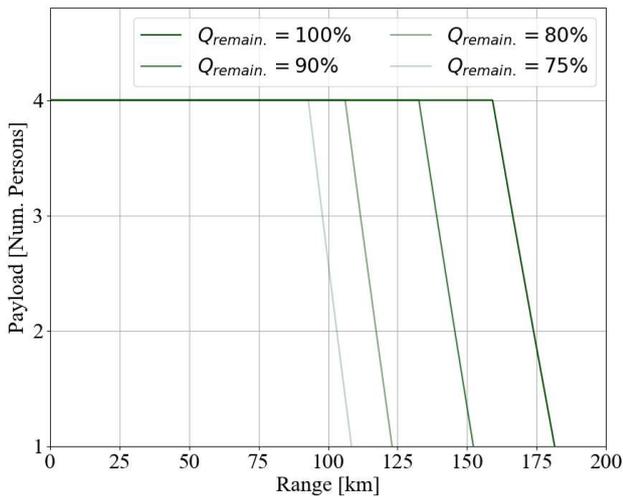

**Fig. 6** Capacity fade for tiltrotor aircraft with enhanced technology

The short battery life might create significant challenges for the successful operation of tiltrotor aircraft in the near-term scenario. Thus, the tiltrotor is also evaluated for the long-term scenario. In the far-term scenario, the aircraft technology has improved, while the battery degradation remains unchanged. The capacity fade for the tiltrotor aircraft with enhanced technology is displayed in Fig. 6 for eleven daily flights. The varying colour intensities represent the payload in terms of persons (passengers and a potential pilot). The darkest tone represents a constant transportation of four persons, while the lightest blue displays the fade for one person. After 610 to 680 flights, 80% of the original capacity is reached. That battery lifetime is an improvement of a quarter compared to the short-term scenario. Additionally, improved technology decreases the energy required for flight, requiring 71% of the initial battery capacity. Hence, the design allows the usage of further degraded batteries. Approx. 940 to 1,050 flights could be carried out until the battery reaches 75% of its initial capacity, which would extend the battery lifetime by about a half.

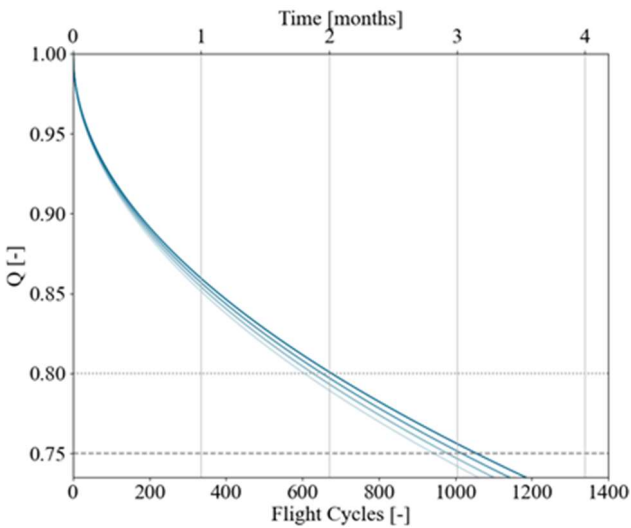

**Fig. 7** Payload-range-diagram for a degrading battery

Fig. 7 shows the corresponding payload-range-diagram for the long-term tiltrotor concept. For example, the range initially is 160 kilometres with four persons on board, but decreases to 107 kilometres as the remaining capacity reaches 80%. The significant reductions in range and payload due to battery degradation pose challenges for aircraft assignment and planning. The increased complexity of restricted operations, including maintenance considerations, has been initially addressed [66, 67].

The realisation of UAM also depends on a competitive price [8] which is countered by a short battery replacement interval. A short-distance multirotor with a near-term design could reach 1,500 to 2,500 flight cycles per battery. The near-term implementation of tiltrotor aircraft would require replacement interval of shorter than one month for a highly utilized aircraft. For a far-term scenario, the improved aircraft technology could extend the number of battery cycles to 650 to 1000, depending on the precise ageing characteristics and payload carried. Lastly, the fading battery capacity has an impact on the dispatchability, thus creating further challenges and potential bottlenecks in flight scheduling, which must be considered in an evaluation for a potential UAM operation.

### 4.1.3 Onboard systems

One of the main challenges preventing VTOL vehicle with passengers on board being already operated today is the unproven safety and reliability of those concepts for commercial passenger operation. Within the last four years the EASA has established rules, named the Special Condition for small-category VTOL aircraft, as well as corresponding means of compliance [21, 68]. Vehicles that are transporting passengers over congested areas fall into the category enhanced of the EASA-SC-VTOL [21]. The safety objectives stated therein for the category enhanced aircraft set high standards similar to commercial aviation. For example, the vehicle must be able for a continued safe flight and landing even if any single system failure occurs and the failure probability for a catastrophic failure must be less than $10^{-9}$ failures per flight hour. As a reliable propulsion system is crucial for a safe VTOL operation, the design of the vehicle's propulsion system takes over an important role. So far, there has only been little research focus on analyzing the propulsion system reliability and its effects on the safe vehicle operation. However, the research that has already been conducted, indicates that the EASA safety objectives are especially difficult to meet for wingless multirotor concept vehicles [69, 70].

Resulting from these challenging safety regulations, a conceptual design process was developed that aims at facilitating the development of a safe propulsion system architecture for multirotor eVTOL vehicles for successful certification. This methodological approach was then applied to design, size and validate the propulsion system for the HorizonUAM multirotor eVTOL concept for an intra-city operation. As a result, a safe battery-electric propulsion system could be designed that is expected to fulfill the EASA SC-VTOL safety requirements. The applied design method is divided in five steps. Within step one, the concept of operation was defined, which includes defining the flight mission and payload requirements. Based on these requirements the vehicle configuration was preselected and the powertrain technology to be used was defined. Within step two several further requirements were developed which are based on the required controllability, the handling quality and allowed noise emission. Within the third step, the propulsion system was defined which can be segregated into defining the flight control system, the power and drive system, the electrical system and the thermal management system considering the previously established requirements. This propulsion system concept was then refined within the safety analysis by taking into account the guidelines ARP4754A [71] as well as the ARP4761 [72]. Thereafter, the vehicle concept was sized as well as validated within the vehicle sizing and simulation step. The system architecture refinement process is usually an iterative process between the concept definition, the safety analysis and the succeeding sizing step and is being conducted until the safety requirements of EASA SC-VTOL can be met. A detailed description of the design process and its application for an exemplary multirotor vehicle and a corresponding propulsion architecture can be found in [73].

The conceptual design method primarily adds value towards improving the design process, with the aim of increasing the reliability of the vehicle and its future operation as well as its safety. It can prevent future high vehicle development costs due to late design adjustments and also help to complete certification procedures quickly and successfully. Furthermore, by increasing the reliability of the design of safe vehicles, the passenger's acceptance can also be expected to be positively influenced.

**4.1.4 Cabin Design**

Within the Horizon UAM project, various factors were considered in the examination and development of the overall UAM system. One of these factors is the cabin. In the design process of the UAM cabin concept, Design Thinking (DT) was applied [74]. DT is a human-centered design approach aimed at generating innovative concepts based on a deep understanding of human needs. This offers a creative and effective approach that emphasizes users' emotions, enabling more effective solutions for various stakeholder requirements. DT was applied to design cabins for the airport shuttle and the intra city use case and led to important advantages for the development process of the air taxi system as a whole in terms of costs, acceptance and safety, flexibility and convenience. The key findings are:

Initial estimates based on commercial aircraft interior masses suggest a possible total weight of 766 kg for the airport shuttle concept. By applying DT and by using lightweight components, a minimalist seat design and a simplified luggage storage solution, the cabin mass for an air taxi concept, could be reduced substantially to an estimated range of 380 kg and 579 kg. The weight reduction can have a positive impact on the sizing of the battery and the required power for air taxi transportation.

In addition, the user-centered design process enables further potential for cost savings. The early feedback process as well as the direct translation of main requirements into a concept can help to avoid cost-intensive adjustments in the late development process.

The direct involvement of different user groups in the design process offers great potential for creating awareness about UAM, disseminating information, and addressing concerns and fears. Particularly during a Focus Group Study, an increased acceptance of this novel mode of transportation was observed. The passengers influence on the design can have a positive impact on acceptance and the perception of safety, leading to a higher willingness to use air taxis among the general population. Moreover, by addressing fears, desires, and concerns directly and incorporating them into the design concept in collaboration with user groups, the development process of autonomously operated air taxis can lead to increased acceptance in the next step.

By combining the airport shuttle and intracity use cases, a multifunctional cabin concept has been developed covering two different application scenarios. The versatility of the cabin concept leads to lower development costs compared to designing separate cabins for each use case. At the same time, the recognition value of the concept increases when used in multiple scenarios, which can positively impact the perceived safety and usability of the cabin features. With its minimalist and interchangeable cockpit design, the cabin can also be changed into a fully autonomous scenario with four passengers in the future.

In addition to the improved seating comfort, optimized storage compartments, minimalist design and customizable privacy features, the cabin design incorporates various comfort parameters based on the feedback from potential user groups. The deliberate combination of minimalist and easily understandable functions with futuristic and complex design elements enhances the overall comfort and user experience.

**4.1.5 Cabin Simulation**

A vital aspect for the economic success of a new technology is that potential users accept the technology. User-centered design aims at introducing the perspective and needs of users of a systems as early as possible in the design process. For that reason, a human-in-the-loop (HITL) simulator representing an air taxi cabin was set up within the project HorizonUAM. The mixed reality (MR) Air Taxi Simulator is a fixed-base simulator comprising a life-size cabin mockup and a video-see-through head-mounted display. The cabin can be augmented with virtual elements but also real people interacting with the participants can be included. Mixing reality with virtual content enables a test person to experience a realistic simulation of an air taxi flight without being entirely decoupled from reality like in conventional virtual reality. A detailed description of the simulation implementation and results regarding the immersion and fidelity of the simulator can be found in [75].

As described in [76], an extensive human factors study with 30 participants using the MR Air Taxi Simulator assessed the factors information need, influence of a flight attendant on board and a re-routing of the flight and their effect on perceived well-being and interaction. The results show that the presence of a flight attendant had no statistically significant influence on the well-being of participants and 16 out of 30 participants stated that an attendant on board is not necessary. Nevertheless, eight found it reasonable for the introduction phase and nine remarked an increase in perceived safety due to the flight attendant. Furthermore, the results show that with an attendant on board, the re-routing scenario was rated better compared to the scenario without an additional crew member on board. With respect to information needs, the three top-most relevant information were travel time, changes of flight route due to obstacles or other traffic, and flight route.

The results show that mixed-reality simulations are a fruitful tool for the investigation of acceptance aspects in order to further shape interaction concepts between passengers and highly automated transportation systems. Using the MR Air Taxi Simulator for an empirical study helped to clarify requirements of end users with regards to crew members on board and information needs. The approach is useful to get qualified user requirements for an acceptable concept and technology and offers a more cost-effective manner to make design decisions in the early stage of the development process.

## 4.2. Infrastructure

UAM ground infrastructure is important for facilitating the safe and efficient integration of UAM operations in the air and on the ground. Vertidromes are specialized (e)VTOL infrastructure components, that serve as essential hubs for future air taxis in urban environments and where users can enter and exit the UAM network. Due to the complex operating environment (urban, often controlled airspace) and high ground and air risks, urban vertidrome networks contribute significantly to the realization of UAM by shaping the air taxi fleet composition, air traffic management, and overall efficiency and throughput capabilities. Therefore, it is crucial to integrate vertidrome research into current UAM development activities.

Due to the lack of full-scale air taxis and fully functional vertidromes capable of accommodating high-density air taxi traffic, both individual vertidrome and vertidrome-network research and planning activities are currently conducted in simulation. Several simulation tools have been developed for this purpose [77]. To study airside and ground traffic flows, a Vertidrome fast-time micro-level simulation model has been developed, which can be used at the strategic planning level considering current vertidrome utilization, occupancy and traffic flows. It allows to determine the operational impact of non-nominal conditions, considering different scheduling methods, operational precision, weather impact, and vertidrome layouts, and thus to evaluate operational capabilities and assess specific vertidrome locations. Within the vertidrome research topic, the system boundaries have been extended to the vertidrome network perspective by developing fast-time agent-based simulation tools for vertidrome airspace network management and vertidrome network optimization tasks [78]. In addition, a real-time human-in-the-loop simulation setup was developed to evaluate the impact of introducing air taxi operations into the controlled airspace and airport environment on air traffic controller workload. First real-time scaled flight tests in a scaled urban environment were also conducted in our temporary modular test facility at DLR's Experimental Test Center for Unmanned Aircraft Systems in Cochstedt, Germany.

Detailed results showed, among other things, that there is significant potential to reduce fleet and ground infrastructure, depending on battery recharge time [79]. The potential for travel time savings could also be demonstrated by choosing the optimal airspace management approach [80]. Regarding the integration of UAM traffic into the air traffic control (ATC) work processes, the real-time simulations showed an increase in workload and situation awareness, suggesting the addition of a working position dedicated to the handling of aircraft flying under visual flight rules (VFR) and UAM traffic in case of increasing UAM traffic [78].

As summarized in [reference CEAS Paper HAP 4], the vertidrome research not only provided valuable detailed results, but also highlighted the complexity of integrating vertidromes into the urban environment and into the pre-existing aviation ecosystem from a procedural and collaborative perspective. There is no one-size-fits-all solution. Therefore, a UAM tailored Vertidrome Airside Level of Service (VALoS) rating concept was developed in order to evaluate the suitability of a specific vertidrome design and location based on those stakeholders relevant for each specific use case [42, 81, 82]. Scalable operational concepts, efficient air traffic management, and optimized allocation and use of network resources are critical to ensure that vertidromes evolve sustainably and successfully along with the growing demand for UAM. The developed tools and methods support the design and evaluation of individual vertidromes in the strategic planning phase, the network of vertidromes and the interaction with airspace network management, fleet design and operations. They are a necessary prerequisite for the simulation of the entire UAM system with the aim of optimally tuning the individual components of UAM in order to design an optimized system adapted to the local conditions.

## 4.3. Operations

Regarding the operation of air taxis, HorizonUAM considered aspects of airspace integration, autonomy, navigation, communication, and associated operating costs.

### 4.3.1 U-Space

Air traffic management (ATM) is a mandatory asset for the safe operation of any air vehicle in controlled airspace. This is especially true for potentially high numbers of future airspace users as unmanned aerial systems (UAS) or air taxis [83]. While piloted air taxis could be operated similarly to helicopters today (e.g. under visual or instrument flight rules, VFR/IFR), remotely piloted or autonomous air taxis will require new ATM solutions to be operated in high numbers.

The European Commission has implemented its U-space regulation [84] as framework for Unmanned Aircraft System Traffic Management (UTM). In general, U-space refers to a set of technologies, procedures, and regulations that enable safe and efficient operations of UAS in low-altitude airspace. It encompasses various aspects of operations, including registration, flight planning, communication, surveillance, and conflict resolution. The concept aims to ensure the integration of UAS into the existing aviation ecosystem, promoting safety, security, and scalability.

In HorizonUAM prototypical U-space services were used to demonstrate vertidrome management tasks [85]. A central U-space cloud service was simulated through a local messaging server using the protocol MQTT (Message Queuing Telemetry Transport). A prototypical vertidrome management tool was created to demonstrate the scheduling and sequencing of air taxi flights. The vertidrome manager was fully integrated within the U-space environment and received real-time information on flight plans, including requests for start and landing and emergency notifications. Additional information coming from other U-space services (e.g. weather information) could be accessed on request. The integration was successfully demonstrated in the erected model city at Cochstedt, Germany, as a scaled flight test environment with multicopters (<15 kg) representing passenger carrying air taxis [85].

The implementation of U-space has direct effects on the feasibility and success of the UAM eco-system. First of all, a high degree of automation on the vehicle side but also on the ATM/UTM side leads to less personnel costs (pilots and air traffic controllers). Secondly, U-space is needed to manage and increase airspace capacity. This will result in higher system efficiency and in higher profit for the operators. Thirdly, efficient (on-demand) flight planning can increase the speed of transport and the convenience for the user. Furthermore, flight path planning and optimization and also the creation of no-fly-zones through U-space can be used to minimize negative effects on the ground such as noise or visual pollution. Finally, U-space ensures safe operation of UAS and air taxis which is essential for the societal acceptance.

### 4.3.2 Safe autonomy and safeguarding machine learning components

Safety is arguably the most important factor for any travel modality. However, to achieve truly affordable and ubiquitous air travel, autonomy is also a key component. With increased autonomy, the operating costs can be reduced. The automation can reduce the training costs of the pilot, for both onboard and remote pilots. By operating without an onboard pilot, the weight of the vehicle can be considerably reduced. This reduces the energy required per mile and thus increases the range. Alternatively, one more passenger or additional luggage can be accommodated. Finally, autonomy enables the scalability and seamless operation of air taxi services to large numbers. This would not be possible with the requirement of aviation pilots for each vehicle and manual coordination and management.

One specific feature explored in the HorizonUAM project was visual person detection. This detection can be utilized for automating the landing approach, so that no person is in danger at a vertidrome. It would be also possible for delivery drones to drop packages in safe distance to persons. Finally, it would be possible to reduce risks in flight by avoiding detected people in the flight path.

Overall, this is a safety task that can otherwise only be performed by a human pilot. Recently, machine learning (ML) has made significant progress in all domains. For this, a ML algorithm was developed and flight tests were performed to gather training and validation data. However, the focus of the research is not the development of an ML algorithm, but to research safety aspects of ML in the context of the aviation domain. There are higher requirements of functional safety in the aviation domain. The use of ML is problematic, since it is considered a black box from a verification point of view. As a result, there is a demand for research in the area of ML safety. Our research started with a literature review on the topic of ML safety as well as the existing standards and regulation for the use of ML in aviation. A research focus was on analyzing the environmental conditions of the operation. In the context of automation systems, this is called operational design domain (ODD). The idea is to check if the specific conditions during operation are consistent with the operational conditions that are expected and that were met during development and training of the ML component.

Furthermore, a safe operation monitor was developed and extended to cover aspects of the ODD to safeguard the ML component as well as the operation. Further details on the topic of safe autonomy and safeguarding ML components can be found in [86].

### 4.3.3 Multisensor navigation

The performance of the navigation system is a key factor to the safety of operations in urban air mobility applications. Vehicles are expected to operate in urban environments, especially during the take-off and landing phases at vertiports. In addition, the costs and size requirements of both vehicles and ground infrastructure are more stringent compared to current civil aviation applications. This leads to new technical challenges to the navigation systems. These challenges may even become increasingly critical as the market evolves and the UAM traffic densities increase in the future. In order to simultaneously achieve high accuracy, high availability and high integrity in urban environments, multisensor navigation solutions are required. However, there are still standardization gaps in the certification of safe multisensor navigation systems. In the HorizonUAM project, we proposed a preliminary architecture design for UAM navigation, and developed innovative methods to quantify the integrity of different subsystems, given the fact that standards are missing for some sensors and for the UAM operation environments. Proof-of-concept demonstrations for our system design were carried out in measurement campaigns using multirotors. The details of the system design and tests can be found in [87].

### 4.3.4 Robust and efficient communication

For the realization of future urban air mobility, reliable information exchange based on robust and efficient communication between all airspace participants will be one of the key factors to ensure safe operations. Due to the high density of piloted and new remotely piloted and autonomous aircraft, air traffic management in urban airspace will be fundamentally different from today. Unmanned Aircraft System Traffic Management (UTM) will rely on pre-planned and conflict-free trajectories, continuous monitoring, and existing communications infrastructure to connect drones to the UTM. However, to mitigate collisions and increase overall reliability, unmanned aircraft still lack a redundant, higher-level safety net to coordinate and monitor traffic, as is common in today's civil aviation. In addition, direct and fast information exchange based on ad hoc communication is needed to cope with the very short reaction times required to avoid collisions and the high traffic density. In particular, the urban environment is very challenging from a physical layer point of view, with rich multipath signal propagation as well as shadowing and diffraction events when flying close to surrounding objects such as tall buildings. Therefore, we are developing a drone-to-drone (D2D) communication and surveillance system, called DroneCAST, which is specifically tailored to the requirements of a future urban airspace and will be part of a multi-link approach. In order to evaluate the reliability and performance of transmission systems, an accurate channel model that takes into account the specific underlying propagation characteristics is essential. As a preliminary step, we have conducted a wideband channel measurement campaign with hexacopters to accurately measure the D2D propagation characteristics in an urban environment. During the HorizonUAM project, we have proposed a novel channel model architecture to model the D2D communication channel for urban environments, which considers the propagation elements and effects identified from our measurements, and shall serve as a basis to easily incorporate further statistics from any related measurements. Furthermore, we presented a multi-link approach with a focus on an ad-hoc communication concept that will help to reduce the probability of mid-air collisions and thus increase social acceptance of urban air mobility. As a first step towards an implementation, we equipped two drones with hardware prototypes of the experimental communication system and performed several flights around the model city to evaluate the performance of the hardware and to demonstrate different applications that will rely on robust and efficient communication. A general discussion on robust communication for urban air mobility and our steps towards DroneCAST can be found in [88].

**4.3.5 Cost modelling**

As long as passengers decide on their mode of transport based on their willingness to pay, the ticket price remains one of the most critical metrics in transportation. In the long term, a ticket price is directly linked to the operating costs of an UAM operator. Thus, a model for the estimation and forecast of direct operating costs becomes one of the key models that a UAM operator needs to develop.

At this moment, many components of the UAM system that contribute to the direct operating cost are not yet known. Therefore, a model for the estimation of the direct operating costs was developed [45]. Where applicable, models of cost components from manned aviation were adapted to estimate UAM parameters. Where this was not feasible, assumptions with an accepted and predictable level of uncertainty were incorporated into the operating cost model.

Direct operating costs were evaluated for different use cases. Depending on their requirements, each use case can be served by various vehicle configurations. Using this information, the cost model calculated different cost scenarios based on a set of input parameters, considering uncertainties. By providing optimistic and conservative scenarios for the direct operating cost as a share of the total cost, indirect costs were also estimated.

The analysis showed that the nature of the direct operating costs depends on the use case and its vehicles. The fare per kilometer highly depends on the considered mission length. The direct operating costs for the intra-city, airport shuttle, and inter-city use cases were estimated. Fares would need to be at least between 4.10€ and 8.50€ per km in order to cover total operating costs and to achieve a 10% profit margin for the first two use cases, respectively. Minimum fares for the intercity use case are estimated to be between 1.00€ and 1.40€ per km. All estimates are based on a hypothetic load factor of 1, which illustrates the challenge of making urban air mobility affordable through low operating costs.

The operating cost and its resulting ticket fare represent some of the most critical metrics for UAM operators. The results presented demonstrate that both mission design and vehicle configuration have a significant impact on the total operating cost per flight. UAM operators need to carefully consider different vehicle configurations, particularly when demand is low and larger vehicles are operating with lower load factors, as reducing operating costs and thus fares is critical to generating sufficient demand to be profitable [45]. Another option could be autonomous vehicles, leading to a reduction in operating costs per seat. While the affordability could increase with the absence of a human pilot, passenger acceptance could potentially decrease. Further work on cost estimation is needed as not all components of UAM have been fully elaborated yet, leading to operating costs calculated with a level of uncertainty that may challenge the guarantee of a profitable airline business. In the future, the current cost model can be expanded with additional insights of UAM components.

### 4.4. Overall system and acceptance

This section briefly describes the results of a market potential analysis and public acceptance studies. Finally, the system-of-systems simulation developed within the HorizonUAM project is presented.

**4.4.1 Market potential analysis**

A preliminary estimate of the potential global demand for UAM, the associated aircraft movements and the required vehicles is essential for the UAM industry for their long-term planning, but also of interest to other stakeholders such as governments and transportation planners to develop appropriate strategies and actions to implement UAM.

As part of the HorizonUAM project, a forecasting methodology is proposed and has been set up to provide estimates of the potential global UAM demand for intra-city air taxi services [89]. The concept is based on a city-centric approach that uses a limited number of city parameters to estimate the total transport demand for each city. A simplified multinomial logit model is used to determine the probability that travelers will choose the air taxi for their individual trips within a city, using travel time and travel costs of each mode as input parameters. Based on the resulting UAM demand, cities with potential for UAM services can be identified. By summing up all city-specific results, an estimate of global UAM demand is provided. Variation of major characteristics of the UAM transportation system allows different scenarios to be developed and analyzed.

Sensitivity analyses were conducted to investigate the impact of vertiport density as well as ticket prices on UAM demand. As expected, UAM demand is highest when air taxi prices are low. However, it is remarkable how strongly demand declines as air taxi prices increase. More vertiports are only economically justified if they generate sufficient UAM demand so that the additional revenues surpass the associated costs. Considering different development paths for air taxi prices and vertiport density, four potential market development scenarios were outlined. The most favorable outcome combines low air taxi prices with high vertiport density, which could result in a market potential for UAM of 19 million daily trips in over 200 cities globally by 2050. Among these cities are international metropole regions such as London, Tokyo, or New York but also major German regions like the Rhine-Ruhr region, Berlin, Munich or Hamburg.

The results emphasize the significant impact of low ticket prices and the necessity of high vertiport density in order to drive UAM demand. This highlights the need to carefully optimize system components in order to minimize costs and increase the quality of UAM services, which will significantly contribute to the economic feasibility and successful implementation of UAM systems.

**4.4.2 Public Acceptance**

In order for UAM to achieve public acceptance and become reality, citizens' concerns and attitudes need to be considered. Otherwise, new technological concepts would be developed without future users as well as affected residents in mind. This would pose a huge risk to the overall UAM system.

One challenge for research on the acceptance of UAM is that UAM is not yet part of citizens daily life. A way of conquering this challenge within the HorizonUAM project consisted in using Virtual Reality (VR) to give citizens simulated experience, either as pedestrians [90] or as air taxi passengers [76]. In addition, potential users were involved throughout the process of designing future air taxi cabins [74].

In order to reveal perceived risks and benefits of UAM in a way that is representative for the entire population, a large-scale telephone survey was conducted. Computer-assisted telephone interviews with 1001 respondents were realized in 2022 to determine current opinions of the German population on civilian drones and air taxis (for a detailed report, see [91]). Overall, civil drones tended to be evaluated rather positively, while no such trend was evident for air taxis in the survey. Answers regarding the attitude towards air taxis ranged from very negative to very positive. The majority of German residents were concerned about the potential misuse of civil drones for criminal purposes as well as the violation of privacy. Provided that citizens' needs are taken seriously, safety and security aspects as well as privacy, therefore, have to be given special consideration in future system design. Moreover, the willingness to use air taxis in the future was most pronounced for use cases involving rural areas. This indicates that the overall system should not strictly be limited to urban airspace [91].

Another way to actively involve citizens is to foster their participation in the assessment of potential impacts of UAM. For instance, the public could be involved in measuring the noise of drone traffic in regional airspace. For this purpose, the smartphone app DroNoise was developed according to the concept of Eißfeldt [92]. DroNoise is suitable for both Android and iOS. Acoustic measurements are taken using either the smartphone's internal or a connectable external microphone. When measuring noise, the maximum value of the A-frequency-weighted and F-time-weighted sound level is recorded (LAFmax). In the course of measurement, users are also asked for their subjective noise annoyance. The history of measurements can be displayed and there is a map that can be used to track the user's location.

DroNoise was calibrated and tested during live flight demonstrations with civil drones at DLR's National Experimental Test Center for Unmanned Aircraft Systems in Cochstedt, Germany. It was demonstrated that DroNoise is executable and suitable for measuring and evaluating drone noise. As a next step, it is planned to test the app on a larger sample for its practicability and applicability. In the long term, DroNoise will be distributed in public app stores to provide access to the entire population. In this way, noise exposure maps could be calculated, which, in turn, would allow traffic management systems to distribute UAM noise pollution among the population in an objective and transparent manner. While the overall UAM system needs to consider the public's well-being and health, DroNoise can provide a basis for addressing this challenge (see also [92]).

### 4.4.3 System-of-systems simulation

Since UAM is a complex system of systems (SoS), with various technical, operational, regulatory, and social components interacting, a holistic view is essential. The complexity results from the integration of the constituent systems such as aircraft, infrastructure, air traffic management and flight operations into the urban transportation system.

Simulations are often employed to optimize specific UAM components. In doing so, they tend to emphasize isolated constituent systems. However, the focus on a single system usually comes at the expense of modeling fidelity of the other constituent systems, which is challenging in a UAM SoS that relies on close integration of the systems. As a high level of interconnectivity is required for the functioning of UAM, it is critical to consider its other constituent systems when designing a UAM system.

Since UAM is currently subject to significant uncertainty due to its early stage of development, integrated modeling of the UAM system of systems can serve as a powerful tool for understanding and reducing uncertainty. In doing so, uncertainties associated with any one of the systems can be evaluated in considering their impact on other constituent systems and the overall SoS.

In order to understand and evaluate the systems of a UAM SoS and their interdependencies, a collaborative agent-based simulation of urban air mobility was developed within the HorizonUAM project. In this context, models for the airside operation of the vertidrome [82], urban airspace management, passenger demand and mode choice estimation, vehicle operator costs and revenues [45], vehicle allocation [79], fleet management based on vehicle design performance, and mission planning were integrated into an agent-based simulation of urban air mobility [93]. The developed collaborative simulation follows a concept of operations as defined within the project, and models UAM from flight request to flight fulfillment over the course of the day for thousands of potential passengers. The collaborative simulation is capable of simulating any combination of vertiport placements, vehicle concepts, fleet combinations, multiple airspace management techniques, vertiport layouts, and demand patterns. Although the use case of focus in developing the simulation was an intracity use case centered in Hamburg, the use case or region can be flexibly changed given the required inputs.

The flight request processing logic of the collaborative simulation is as follows: for each flight request received, it is first assessed whether the request can be allocated to an existing scheduled flight. If not possible under the defined criteria, a new flight must be scheduled. The flight scheduling process consists of first selecting a vehicle from the fleet considering its position, available energy and availability; computing departure and arrival slots and a deconflicted flight trajectory. The flight information and ticket price alongside alternate mode options are then passed on to the mode choice model for the final passenger decision. If UAM is chosen by the passenger, their seat on the flight is fixed, and if not, the seat (and potentially flight) is not fixed.

The integration of the individual modules (or constituent systems) into the agent-based simulation was achieved through the use of Remote Component Environment (RCE) [94, 95], which served the crucial function of seamlessly connecting the models hosted across several cities and institutes. By modelling the key subsystems involved in UAM, and simulating over the course of hours/days, the influence of subsystem level parameters on the overall network can be analyzed. The integration and orchestration of closed models located at different sites within an agent-based simulation, thereby constituting a collaborative system of systems simulation, was demonstrated within this project and the process of implementation, performance optimization and proof of concept is presented in [96].

Such a comprehensive approach allows to capture the multi-layered interactions and interdependencies within the UAM ecosystem. By taking a holistic view is it possible to understand and assess the challenges, opportunities and potential of UAM to support the effective integration and sustainable success of urban air mobility as new mode of transportation. Furthermore, such an approach can also allow the combined optimization of the individual constituent systems and the overall system of systems.

## 5. SUMMARY AND CONCLUSION

For UAM to become reality, it needs to be socially accepted in the broadest sense. Here, acceptance includes aspects of socio-political acceptance, community acceptance and market acceptance. Thus, UAM has to meet the expectations of all UAM stakeholders such as potential users, industry, governments, public institutions, regulators and indirectly affected third parties, reconciling their own motivations, expectations and concerns. For this, the UAM system needs to be designed in a way that promotes benefits and minimizes risks, so that UAM as part of the future urban transportation system is safe, affordable, accessible, environmentally friendly, economically viable and finally sustainable. In order to optimize the UAM system with regard to the desired characteristics, it has to be considered as a highly complex system-of-systems, where the overall system performance results from the interaction of its components. Not only the individual system components should be optimized. A holistic view on the overall UAM system considering the interaction between the individual components is mandatory.

The DLR project HorizonUAM has contributed to both aspects of UAM research, the development of individual components as well as their harmonization for use in an optimized overall system, by combining research on UAM vehicles, related infrastructure, operation of UAM services, and public acceptance of future urban air transport. In particular, the complexity of Urban Air Mobility with its interdependencies has been addressed by the project.

Key results of HorizonUAM are:

- A simplified model-based estimation taking into account size and wealth of cities show that there could be high market potential for UAM in more than 200 cities worldwide. Among these cities are international metropole regions such as London, Tokyo, or New York but also major German regions like the Rhine-Ruhr region, Berlin, Munich or Hamburg. Future research should also include regional operations.

- Passengers' and pedestrians' perspective on UAM and attitude of the German population were evaluated in order to assess acceptance of UAM by users and non-users. The German population is currently undecided about UAM. Answers regarding the attitude towards air taxis ranged from very negative to very positive. Acceptance was highest for use cases including rural areas. Further research should also be dedicated to the design of human-machine-interfaces for passenger of vehicles without pilot on board.

- Models from different domains were integrated to a system-of-system simulation suitable for the analysis of UAM systems. Here, tools for vertidrome assessment, vertidrome airspace network and fleet management optimization go in hand with tools for demand estimation, cost and revenue estimation, and mode choice and vehicle design.

- Vehicle concepts suitable for intra-city and sub-urban use cases were designed. A quadrotor configuration was suggested for ranges up to 50 km and a tiltrotor configuration for longer ranges up to 100 km.

- A Drone Communications and Surveillance Technology (DroneCAST) was developed to enable drone-to-drone communications for collision avoidance in urban airspace.

- A Safe Operation Monitor was developed to address certification aspects of machine learning for improving the reliability and trustworthiness of artificial intelligence functions.

- A Vertidrome Level of Service framework was established that is suitable to rate vertidrome airside operations unifying multidisciplinary stakeholders.

- Vertidrome integration into airport environment and thus controlled airspace was assessed with air traffic controllers and revealed the limitations in runway capacity and acceptable workload. An exclusive air taxi working position and further automation of processes is recommended. The establishment of scalable U-space services for traffic management of drones as well as passenger carrying air taxis is key to solve these issues.

- A modular model city was erected as a scaled environment for demonstration of airspace integration, vertidrome management, artificial intelligence functions, and urban communication and navigation. This model city is flexible in design and available for future UAM evaluations and demonstrations.

The results of HorizonUAM indicate that UAM could become technically feasible in the near future. However, the following key challenges need to be addressed before UAM can be widely implemented:

- Profitability: For their widespread adoption and acceptance by users, manufacturers and investors, it is essential to ensure that air taxi services are economically viable even with low ticket prices. This requires to minimize direct and indirect operating costs of the UAM transportation system. Suitable business models need to evolve for UAM to become more than a niche market. The evolving regulatory framework for UAM needs to be matured and harmonized internationally in order to ensure safety, security, and environmental sustainability but also scalability in order to make UAM financially feasible.

- Complexity of the UAM system: Managing complexity and filling existing knowledge gaps to remove uncertainties is necessary to achieve high efficiency of the UAM system. This results in a complicated distribution of responsibilities among UAM stakeholders including users, industry, governments, public institutions, regulators and communities. All of these stakeholders must work together to shape the transportation system of the future. In particular, the interactions of the individual UAM system components, its interdependencies and the effects on the feasibility of the overall transportation system need to be further investigated to develop an economically viable and scalable UAM system that maximizes the benefits not only for the users but the society in general.

- Social acceptance and in particular community acceptance: Acceptance may be one of the critical factors in UAM implementation in many societies. Appropriate measures will need to be taken to address the key concerns of noise, actual and perceived safety and security, high energy consumption, visual pollution and land use. In order to offer seamless transportation, the integration of UAM into existing transportation networks is essential and can improve the efficiency of the overall transportation system with benefits for users and society. It is of highest importance to keep the general public informed about urban air mobility and its implications. Communities have to be actively engaged in the design of a potential future transportation system to make it a success. Therefore, information based on scientific analysis but tailored towards a non-scientific audience should be provided by the UAM community. Real live demonstrations are recommended to increase the familiarity with UAM in the general public.

In conclusion, it has been shown that UAM might complement existing transportation systems in the future. Ultimately, it is a matter of the constituent systems working together in a way that the overall system is both economically feasible and socially acceptable.

DLR will continue to work on the idea of Urban Air Mobility. Future research will be extended by considering new multimodal and regional use cases. Thus, the initial urban scope will be extended to evolve form Urban Air Mobility over Advanced Air Mobility to Innovative Air Mobility with the overall goal of integrating drone and air taxi services into existing transportation systems.

## COMPETING INTERESTS